\pdfoutput=1
\documentclass[fleqn,10pt]{wlscirep}
\usepackage[utf8]{inputenc}
\usepackage[T1]{fontenc}

\usepackage{graphicx} 
\usepackage{graphicx,lineno,bm,float,soul,amsfonts,amsmath,siunitx}
\usepackage{xcolor,hyperref}
\usepackage{empheq}
\usepackage{multirow}

\newcommand{\B}[1]{{\bm #1}}

\newcommand{\ds}{\displaystyle}

\title{Connecting Earth and Moon via the L1 Lagrangian point}

\author[1,2]{A. K. de Almeida Jr}
\author[3,2]{V. M. de Oliveira}
\author[1,2]{T. Vaillant}
\author[1]{D. Maia}
\author[2,4]{A.C.M. Correia}
\author[5,1]{D. Barbosa}
\author[6]{L.T.B. Santos}

\affil[1]{CICGE, DGAOT, FCUP, Vila Nova De Gaia, Portugal}
\affil[2]{CFisUC, Departamento de Física, Universidade de Coimbra, 3004-516 Coimbra, Portugal}
\affil[3]{Instituto de Matemática e Estatística, Universidade de São Paulo, 05508-090 São Paulo/SP,Brazil}
\affil[4]{IMCCE, UMR 8028 CNRS, Observatoire de Paris, PSL Université, 77 Avenue Denfert-Rochereau, 75014 Paris, France}
\affil[5]{University of Évora, R. Romão Ramalho 59, 7000 Évora, Portugal}
\affil[6]{Physics of Materials, Polytechnic School, University of Pernambuco, Recife, PE, Brazil}

\begin{abstract}
The renewed global interest in lunar exploration requires new orbital strategies to ensure flight safety which can benefit extended lunar missions and service a plethora of planned instruments in the lunar orbit and surface. We investigate here the equivalent fuel consumption cost to transfer from (to) a given orbit and enter (leave) at any point of an invariant manifold associated with a Lyapunov orbit around the Earth-Moon $L_1$ Lagrangian point using bi-impulsive maneuvers.
Whereas solving this type of transfer is generally computationally expensive, we simulate here tens of millions of transfers orbits, for different times of flight, Jacobi constants and spatial location on the manifold. We are able to reduce computational cost by taking advantage of the efficient procedure given by the Theory of Functional Connections for solving boundary value problems, represented with special constraints created to the purposes of this work.
We develop here the methodology for constructing these transfers, and apply it to find a low-cost transfer from an orbit around the Earth to a stable manifold and another low-cost transfer from an unstable manifold to an orbit around the Moon.
In the end, we obtain an innovative Earth-to-Moon transfer that involves a gravity assist maneuver with the Moon and allows a
long stationed stage at the Lyapunov orbit around $L_1$ which can be used for designing multi-purpose missions for extended periods of time with low fuel costs.
This is paramount to optimize new exploration concepts.
\end{abstract}
\begin{document}

\flushbottom
\maketitle
%
%
\thispagestyle{empty}



\section{Introduction}


Cislunar space is swiftly becoming a nexus of increased scientific exploration interest and geopolitical and business rivalry. Around 250 missions are expected to be launched to the Moon after 2030, including scientific, robotic and human-crewed missions\footnote{\url{https://brycetech.com/reports/report-documents/emerging-industrial-base-lunar-2024/}}. The Moon exploration became an international endeavor with the United States\cite{2022Natur.611..643W}, China\cite{10.1007/978-981-99-8048-2_274,2019NatGe..12..222W}, India\cite{2023Natur.620..927P}, Japan\cite{2023AGUFMSM51B2546Y}, Russia\cite{2024AcAau.222..346T}, South Korea\cite{2024LPICo3040.1779J} and the United Arab Emirates\cite{2024LPICo3063.5070A} all at various stages of lunar exploration missions. While the NASA-led Lunar Gateway program\cite{FULLER2022625} aims to build a multinational space station in orbit around the Moon, the China National Space Administration (CNSA) has a long term goal of establishing a permanent lunar base near the Lunar South Pole\cite{HU2023105623}. The anticipation of growing international collaborations, impending commercial endeavors, and the establishment of ethical frameworks such as the Artemis Accords collectively emphasize the importance of this vast region between Earth and the Moon. 

With new and old spacecraft, aging effects, and inevitably, space debris, a situation similar to the highly populated near-Earth region is being created in the Moon's surroundings.
The much smaller orbital volume around the Moon is mainly due to the perturbation of the gravity of the Earth. It means the orbital space to accommodate ships, relay satellites and planned Moon orbiting space stations together with its growing space debris field will rapidly become crowded if these orbit the Moon at low altitudes above the lunar surface. This situation coupled to the growing appetite to extend Earth orbital space to cislunar space presents challenges to future missions that may require careful or even negotiated allocation of orbit slots to avoid traffic problems.
For instance, Apollo missions orbited just over 100 km in altitude and orbits of lunar satellites within few hundreds kilometers above the lunar surface will become inappropriate options for the numerous planned cislunar/lunar missions.
This situation points to the necessity of considering several innovative orbital approaches, such as using periodic orbits around the Earth-Moon system Lagrangian points.

The Lagrangian points are dynamical equilibria of the Three-Body Problem and they represent interesting solutions for space mission design. In particular, unstable periodic orbits around these Lagrangian points are especially interesting because it takes relatively little energy consumption to keep an object in such solutions. 
For instance, the Lagrangian points of the Earth-Sun system were widely used for space missions.
Thus, many solar observatories like the ESA Solar and Heliospheric Observatory (SOHO)\cite{Fleck2021}, the NOAA Deep Space Climate Observatory (DSCOVR)\cite{6187025}, the ISRO Aditya-L1\cite{2018AdSpR..61..749Y} and the NASA Advanced Composition Explorer (ACE)\cite{1998SSRv...86....1S} were positioned about the first Lagrangian point $L_1$, and many satellites were positioned about the second Lagrangian point $L_2$ as, for example, ESA's Gaia\cite{refId0}, Herschel large infrared telescope\cite{2010A&A...518L...1P}, Planck Surveyor Microwave telescope\cite{2011A&A...536A...1P} and NASA's James Webb Space Telescope\cite{Rigby_2023} \footnote{ \url{https://webb.nasa.gov/content/about/orbit.html}}. Other missions like the CNSA Chang'e 5 lunar orbiter\cite{doi:10.34133/2021/9897105} took a complex  orbit transfer: after dispatching lunar samples back to Earth in 2020, it has sent its transport module from the Moon to L1 where it is permanently stationed to conduct space weather observations.
These are examples of the importance of designing efficient orbit transfers to and from the vicinity of Lagrangian points. 


Due to the fact that the periodic orbits around $L_1$ are unstable, they possess stable and unstable manifolds, sets of solutions converging to these orbits forward or backward in time \cite{koon2000}. Such manifolds form geometrical structures that may provide a natural path between two solutions in the system. For example, there is a big effort in developing numerical methods for efficiently determining crossings between invariant manifolds \cite{henry2023}.
We are interested here in low-energy transit orbits, solutions that exit the Earth's realm and reach the lunar realm via the equilibrium point $L_1$. In \cite{topputo2005earth}, for example, the authors focused on this type of solution to construct Earth-to-Moon transfers using invariant manifolds associated to Lyapunov orbits. They were able to save approximately 100 m/s in $\Delta V$ in the Moon captured arc in comparison to the Hohmann transfer, showing good promise for this type of strategy. Their design however was based only on linear periodic solutions around $L_1$ and their final transfers required multiple impulses. In this work, we consider a Lyapunov orbit far from the linear regime and we use the Theory of Functional Connections (TFC) to construct efficient orbital transfers.

The TFC \cite{leake2022} is a recent theory that has gained increasing attention in areas like Differential Equations \cite{Leake2019,Leake2020,florion2,Wang2024}, Optimization and Optimal Control \cite{Johnston2020,Li2021,Wang2022}, 
Particle physics \cite{Floriononame1,Schiassi2022}, Biology \cite{Xu2020,Daryakenari2024}, and Geodesy \cite{Mortari2022}, among many others fields of the knowledge (such as Fractional operators, Functional Interpolation, Physics-Informed Neural Networks, Nuclear Reactor Dynamics, and Neuroscience). Furthermore, it has been proven as an effective method to design maneuvers in the cislunar space through a model based on the Two Point Boundary Value Problem (TPBVP) \cite{fastTFC}. TFC can be adopted to perform orbit determination and characterize satellites using an inverse problem approach \cite{dealmeida2024}. TFC can also be adopted to perform orbit transfer with the adoption of low-thrust \cite{DEALMEIDA2023102068}, e.g. generated by a solar sail \cite{tfc_solarsail}.
Furthermore, using a suitable change of coordinates \cite{tfcvariables}, the problem can be formulated using other constraints than the ones of the TPBVP \cite{akajtangential}.
We propose the use of the TFC to design transfers from an orbit around the Earth to an orbit around $L_1$ and then to an orbit around the Moon using bi-impulsive maneuvers in each leg of the journey. This is done by formulating the problem based on special constraints aiming to increase numerical convergence and decrease computational costs. We focus on Lyapunov orbits, solutions of the Circular Restricted Three-body Problem (CRTBP) that exist in the vicinity of the collinear Lagrangian points. The priority here is to evaluate such maneuvers and select the best option related to short transfers, for times of flight up to 50 days, as classified in \cite{QI2019180} for transfers from the Earth to lunar libration point orbits. Although longer times of flight can be adopted to decrease the costs, the effects of neglected perturbations, e.g. due to the gravity of the Sun, are accumulated over time, whereas divergence of results is increased.

The models and tools to be adopted in this research are shown in Sec. \ref{sec:models}, where the technique is presented. The results are shown in Sec. \ref{sec:results} for transfers between the Earth and Lyapunov orbits and from this last one to the Moon.


\section{Mathematical models}
\label{sec:models}

The model adopted is given by the Circular Restricted Three-Body Problem (CRTBP), where the Earth and the Moon rotate in circular orbits around their common barycenter. A spacecraft, with negligible mass in comparison with the Earth and the Moon, is in an initial circular orbit around the Earth with an altitude of 167 km. The Lyapunov orbits considered here are located around the $L_1$ equilibrium Lagrangian point and are the intermediary stage of the mission. The final orbit around the Moon is circular with an altitude of 100 km.
The transfer takes place in the plane defined by the motion of the Earth-Moon pair, where the initial orbit around the Earth, the final orbit around the Moon, and the Lyapunov orbits around $L_1$ are also.

\subsection{The Circular Restricted Three-Body Problem}

\begin{figure}
	\centering
	\includegraphics[scale=0.56]{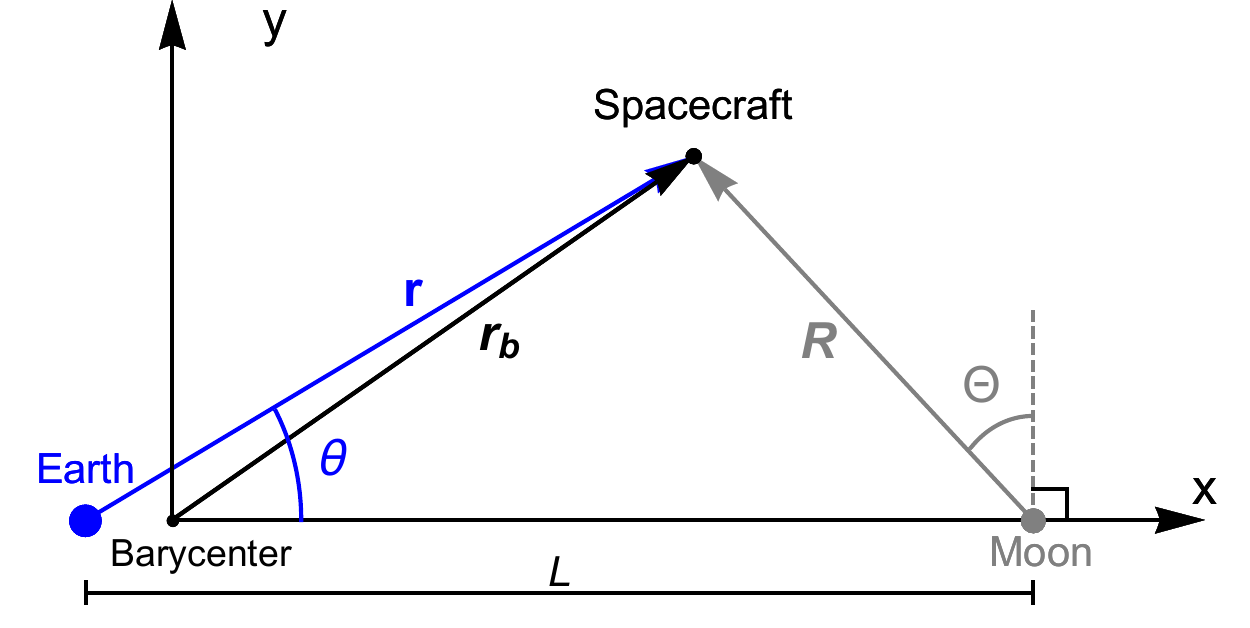}
	\caption{Relative positions of the bodies and variables in the rotating frame of reference.}
	\label{fig:frame}
\end{figure}

Under the Circular Restricted Three-body Problem (CRTBP) model, we assume that both the Earth and the Moon are moving in a circular motion around their barycenter with an angular velocity $\omega$. 
An inertial frame of reference with Cartesian coordinates ($x_i,y_i,z_i$) is defined fixed at the barycenter. A rotating frame of reference with Cartesian coordinates ($x,y,z$) (Fig. \ref{fig:frame}) is defined such that the Earth and the Moon are both fixed in the $x$ axis, while the $z$ axis is in the direction of their angular momenta.
The equation of motion of a satellite with negligible mass in comparison to the Earth and the Moon is then given by \cite{symon}
\begin{equation}
	\label{eq:1}
	\frac{\text{d}^2\B{r}_b}{\text{d} t^2}+2\B{\omega}\times\frac{\text{d}\B{r}_b}{\text{d} t}+\B{\omega }\times \left(\B{\omega }\times \B{r}_b\right)=-\frac{\mu_e}{r^3}\B{r}-\frac{\mu_m}{R^3}\B{R},
\end{equation}
where
$\B{r}_b=(x,y,z)$ is the position of the spacecraft relative to the barycenter, 
$\B{r}$ is the position of the spacecraft with respect to the Earth and with norm $r$, 
$\B{R}$ is the position of the spacecraft with respect to the Moon and with norm $R$,
$\mu_e$ and $\mu_m$ are the gravitational parameters of the Earth and the Moon, respectively,
and $\B{\omega }=\omega \B{k}$ the angular velocity of the rotating frame, 
with $\B{k}$ a unit vector along the $z$ axis and $\omega=\sqrt{(\mu_e+\mu_m)/L^3}$, where $L$ is the total distance separating the two bodies.

The CRTBP admits a constant of motion known as the Jacobi constant $C$ \cite{1967torp.book.....S} given by
\begin{equation}
	\label{eq:C}
	C=\omega^2(x^2+y^2)+2\left(\dfrac{\mu_e}{r}+\dfrac{\mu_m}{R}\right)-\dot{x}^2-\dot{y}^2-\dot{z}^2,
\end{equation}
and that defines a five-dimensional manifold on which the solutions of Eq.~\eqref{eq:1} take place.


\subsubsection{Dynamics around $L_1$: Lyapunov orbits and their invariant manifolds}

The CRTBP has five equilibrium points in the rotating frame known as Lagrange points, all of which are located in the plane $z = 0$. The collinear points ($L_1$, $L_2$, $L_3$) lie on the $x$-axis, while each of the triangular points ($L_4$, $L_5$) form an equilateral triangle with the primaries. The collinear points are unstable for any mass ratio in the system. Linearly, they have a stability of the type saddle $\times$ center $\times$ center \cite{Mingotti2, 2023RMxAA..59...83S}. The two centers are associated with motion inside and outside of the $x$-$y$ plane and indicate that there exist families of planar and vertical periodic orbits, known as Lyapunov orbits \cite{Jorba1, Gomez3, Topputo16}. The factor ``saddle", on the other hand, implies the existence of a set of orbits which converges forward in time to the periodic orbit, called stable manifold, and another set which converges to the periodic orbit backward in time, called unstable manifold \cite{Howell84, Howell88, Gomez2, Topputo16, Priscilla, 2023RMxAA..59...83S}.

In this work, we are going to focus on the planar Lyapunov orbits. It has been shown that the stable and unstable manifolds associated with these orbits create interesting structures in the system's phase space and are closely connected to the system's dynamical properties \cite{vitor2020,vitor2022}. In order to trace these solutions, we first calculate a planar periodic orbit in the linear regime and use a continuation procedure to determine the Lyapunov orbit for any given value of the Jacobi constant $C$. We then discretize said orbit in $N$ points and choose initial conditions in the vicinity of each point. The initial conditions associated with the $i$-th point are chosen in the directions given by the eigenvectors of the Monodromy matrix evaluated at such a point \cite{koon2000}. Finally, we obtain the invariant manifolds of the Lyapunov orbit by evolving the system forward and backward in time using these initial conditions.

\subsubsection{Earth-centered polar coordinates}

The first leg of the transfer leaves the Earth and arrives at the stable manifold transporting the spacecraft to the Lyapunov orbit. 
The equations of motion adopted to solve this transfer are based on the polar coordinates ($r,\theta$) centered on Earth (Fig. \ref{fig:frame}) and are given by \cite{akajtangential}
\begin{equation}\label{eq:motion_earth}
	\begin{aligned}
		\ddot{r} - r \, \dot{\theta}^2 - r \omega^2 - 2 r \, \omega \, \dot{\theta} + \omega^2 d_1 \cos \theta +\dfrac{ \mu_e}{r^2} +  \dfrac{\mu_m(r-L \cos \theta)}{\left(r^2-2 r L \cos \theta +L^2\right)^{3/2}} &=0, \\ 
		r \, \ddot{\theta} + 2 \dot{r} (\dot{\theta} + \omega)  - \omega^2 d_1 \sin \theta + \frac{ \mu_m  L \sin \theta}{\left(r^2-2 r L \cos \theta +L^2\right)^{3/2}} &= 0,   
	\end{aligned}
\end{equation}
where $d_1$ is the distance from the center of the Earth to the barycenter and the angle $\theta$ is measured with the line connecting the Earth and the Moon, being positive in the counter-clockwise direction.
With the use of polar coordinates, the important constraint of a tangential burn for leaving the initial orbit around the Earth is linear, and can be then analytically embedded into Eqs. (\ref{eq:motion_earth}) via TFC \cite{akajtangential}.
This strategy will be detailed in Sec. \ref{sec:strategies}.

\subsubsection{Moon-centered polar coordinates}

Analogously, a polar system of coordinates ($R,\Theta$) centered on the Moon (see Fig. \ref{fig:frame}) is adopted to solve the transfer between the unstable manifolds (leaving the Lyapunov orbits) and the Moon. The equations of motion are
\begin{equation}\label{eq:motion_moon}
	\begin{aligned}
		\ddot{R} - R \, \dot{\Theta}^2 - R \omega^2 - 2 R \, \omega \, \dot{\Theta} + \omega^2 d_2 \sin \Theta +\dfrac{ \mu_m}{R^2} +  \dfrac{\mu_e(R-L \sin \Theta)}{\left(R^2-2 R L \sin\Theta +L^2\right)^{3/2}} &= 0, \\ 
		2 \dot{R} (\dot{\Theta} + \omega) + R \, \ddot{\Theta} + \omega^2 d_2 \cos \Theta -\frac{ \mu_e  L \cos \Theta}{\left(R^2-2 R L \sin\Theta +L^2\right)^{3/2}}  &= 0,
	\end{aligned}
\end{equation}
where $d_2=L-d_1$ is the distance between the Moon and the barycenter and ($\Theta+\pi/2$) is the angle between the position vector and the closed line segment connecting the centers of the Moon and Earth, being positive in the counter-clockwise direction.

\subsection{Transfer strategies}
\label{sec:strategies}

The strategies adopted to solve the transfer problem are described in this section. In Sec. \ref{sec:ELtransfer}, the spacecraft is transferred from the Earth to a Lyapunov orbit using a bi-impulsive maneuver, while the transfer from the Lyapunov orbit to the Moon is described in Sec. \ref{sec:LMtransfer}, using another bi-impulsive maneuver.

\subsubsection{Transfer from Earth to Lyapunov orbit}
\label{sec:ELtransfer}

The spacecraft is initially orbiting Earth in a circular orbit of 167 km altitude, and in polar coordinates, the initial radius is then given by the sum of the altitude of the satellite and of the Earth's radius.
The optimization procedure then determines the point of the orbit where the impulse must be applied to transfer the spacecraft to the stable invariant manifold.
A velocity tangent to the orbit is the most efficient direction to apply the impulse in a two-body problem context \cite{MARCHAL196991}.
A such tangential impulse minimizes the costs of transfers between circular orbits \cite{vallado7}, and it also holds for elliptical orbits if the impulse is applied at the perigee or apogee of the transfer orbit \cite{LAWDEN1962323}.
Such a tangential velocity also improves convergence of numerical optimizers, as the shooting method \cite{topputo2013optimal} or Sequential Quadratic Programming \cite{folta13}.
In this work, we adopt an innovative approach by imposing the tangential velocity restriction - to leave the initial orbit around the Earth - as a constraint, and we use TFC to analytically embed it into the equations of motion of the spacecraft. 
We then extend the approach shown in \cite{akajtangential} combining TFC with changes of coordinates \cite{tfcvariables}.
A tangent velocity constraint is simply represented by a null radial velocity in a polar coordinate system.
The constraints defined in Table \ref{tab:constraints} will be then analytically embedded into the solution via TFC (see Sec. \ref{sec:tfc}).
Hence, any solution obtained later by the numerical optimization procedure will satisfy both the configuration of the transfer (a starting point on the initial orbit around the Earth) and the most efficient direction to apply the impulse. This technique is very important because it eliminates several parameters of the transfer from the numerical optimization problem.
\begin{table}
	\centering
	\begin{tabular}{c|c|c|c|c|c}
		\hline
		Type of transfer&Time&\multicolumn{2}{c|}{Position}&\multicolumn{2}{c}{Velocity}  \\
		& & Radial  & Angular  & Radial  & Angular  \\
		\hline
		\multirow{2}{*}{Part 1: from Earth to stable manifold}&$t=0$ & $r = r_0$& none & $\dot{r} = 0$ & none \\
		&$t=T$ & $r = r_f$ & $\theta = \theta_f$ &none&none\\
		\hline
		\multirow{2}{*}{Part 2: from unstable manifold to Moon}&$t=0$ & $R = R_0$&$\Theta = \Theta_0$ &none&none\\
		&$t=T$ & $R = R_f$& none  &$\dot{R} = 0$&none
		\\ \hline
	\end{tabular}
	\caption{The constraints adopted in the two parts of the motion.}
	\label{tab:constraints}
\end{table}

The dynamics of the spacecraft is described with the equations of motion given by Eq. (\ref{eq:motion_earth}).
The final point of the transfer (coordinates ($r_f,\theta_f$) in Table \ref{tab:constraints}) belongs to the invariant manifold associated to the Lyapunov orbit corresponding to the chosen value of the Jacobi constant.
When the position of the spacecraft coincides with the point ($r_f,\theta_f$), a second impulse is then applied such that its phase-space coordinates (both its position and velocity) are identical to the coordinates of the flow of the manifold. Afterwards, the natural dynamics of the invariant manifold is then used to transport the satellite from this point to the Lyapunov orbit associated to the chosen Jacobi constant. It is important to note that the natural dynamics is an approximation due to the two major gravitational forces, then a control must be used to maintain the spacecraft on its course proposed here and to perform a transfer from the manifold to the Lyapunov orbit using a desired time of flight, according to the objectives of the mission.
This control is not considered in this paper in order to maintain the generality of its results useful to the purpose of preliminary mission analyses.

\subsubsection{Transfer from Lyapunov orbit to Moon}
\label{sec:LMtransfer}

The spacecraft is initially in a Lyapunov orbit associated to the same chosen Jacobi constant it entered. It can be noted that the Lyapunov orbit is unstable, thus a control is required for station keeping purposes, i.e. to maintain it in an orbit around the Lagrangian point $L_1$ for longer periods of time. In this sense, since the dynamics of the orbit is unstable, the spacecraft leaves the Lyapunov orbit through the unstable manifold using a small perturbation given by its control system in the desired direction, whose cost is neglected in this study as explained in the last paragraph of Sec. \ref{sec:ELtransfer}. Afterwards, at some point chosen from the manifold, a first impulse is applied such that a transfer to the Moon is initiated.
Once the spacecraft reaches 100 km altitude around the Moon, another impulsive is applied to circularize the orbit around it. Thus, a bi-impulsive maneuver is adopted to perform the transfer from the manifold (and from the Lyapunov orbit) to the Moon.
The dynamics at this part is described with the equations of motion given by Eq. (\ref{eq:motion_moon}), while the motion is subject to the constraints shown in the Part 2 case in Table \ref{tab:constraints}. It can be noted that, on the contrary of the Part 1 case, the initial position is constrained and the final position is partially constrained: the final radius is constrained, but the final angle $\Theta$ is free and will be determined by the optimization procedure. Similarly to the analysis done in the previous section, the lowest cost is obtained in the case where the impulse to circularize the orbit around the Moon is applied in the tangent direction, satisfying then the condition $\dot{R}(T)=0$ shown in Table \ref{tab:constraints}.

\subsection{Using TFC to embed the constraints}
\label{sec:tfc}

Since polar coordinates are adopted, the constraints shown in Table \ref{tab:constraints} are linear in the dependent variables, allowing TFC to be used to embed them into the solutions \cite{deAlmeida2023}.
The optimum impulsive way to transfer a spacecraft from Earth is applying a $\Delta V$ in the direction tangential to both the initial orbit around the Earth and the transfer trajectory. In this paper, this most efficient way is embedded into the solutions through the constraint $\dot{r}(0)=0$ shown in Table \ref{tab:constraints}. Similarly, the most efficient way to circularize the orbit around the Moon is by applying an impulse in a direction tangential to the trajectory of transfer and the final orbit around the Moon. This condition is embedded in the constraint $\dot{R}(T)=0$ shown in Table \ref{tab:constraints}.

According to \cite{U-ToC}, the \textit{constrained expression} can be obtained from the equation
\begin{equation}\label{eq:CEr}
	\begin{aligned}
		r(t) &= g_r(t) + \eta_0(t, g (t)) s_0(t) + \eta_1(t, g (t)) s_1(t)+ \eta_2(t, g (t)) s_2(t), \\
		R(t) &= g_R(t) + \lambda_0(t, g (t)) s_0(t) + \lambda_1(t, g (t)) s_1(t)+ \lambda_2(t, g (t)) s_2(t),
	\end{aligned}
\end{equation}
where $g_r(t)$ and $g_R(t)$ are free functions associated to the $r$ and $R$ coordinates, respectively, $\eta_k$ and $\lambda_k$, for $k=0,1,2$, are coefficients (constants in time for the purpose of this work, as defined in \cite{U-ToC}), and $s_k(t)$, for $k=0,1,2$, are given support functions \cite{TFC_Book}.
In order to simplify, we choose these support functions as elements of the canonical basis of polynomials of order 2 ($s_0(t)=1$, $s_1(t)=t$, $s_2(t)=t^2$). 

The constraints adopted to design the transfer from the Earth to the stable manifolds for the radial coordinate $r(t)$ and the constraints to transfer from the unstable manifolds to the Moon for the radial coordinate $R(t)$ shown in Table \ref{tab:constraints} can be written as
\begin{equation}\label{eq:rRconst}
	\left\{\begin{aligned}
		r(0) & = r_0, \\ 
		r(T) & = r_f, \\ 
		\dot{r}(0) & = 0,
	\end{aligned}\right.
	\quad
	\left\{\begin{aligned}
		R(0) & = R_0, \\ 
		R(T) & = R_f, \\ 
		\dot{R}(T) & = 0.
	\end{aligned}\right.
\end{equation}
A set of three equations is generated by substituting the variable $r(t)$ shown in Eq.~(\ref{eq:CEr}) into the left part of Eq.~(\ref{eq:rRconst}).
This set of equations is solved for the three coefficients $\eta_k$ for $k=0,1,2$. 
A similar procedure is performed by replacing the variable $R(t)$ shown in Eq.~(\ref{eq:CEr}) into the right side of Eq.~(\ref{eq:rRconst}), whose solution is obtained for $\lambda_k$ for $k=0,1,2$.
These operations ensue the derivation of the following \textit{constrained expressions} for $r(t)$ and $R(t)$
\begin{equation}\label{eq:cer}
	\begin{aligned}
		r(t) =& \,\, r_0 - g_r(0) + g_r(t) - t g^{\prime}_r(0) + t^2 \left[\dfrac{g^{\prime}_r(0)}{T}+\dfrac{-r_0 + r_f + g_r(0) - g_r(T)}{T^2} \right], \\
		R(t) =&  \,\, R_0-g_R(0)+g_R(t) + t\left[g^{\prime}_R(T)+\dfrac{-2R_0+2R_f+2g_R(0)-2g_R(T)}{T}\right] \\
		& +t^2\left[-\dfrac{g^{\prime}_R(T)}{T}+\dfrac{R_0-R_f-g_R(0)+g_R(T)}{T^2}\right],
	\end{aligned}
\end{equation}
where $g_r^\prime(0)$ and $g_R^\prime(T)$ represent the time derivative of the free functions $g_r(t)$ and $g_R(t)$ evaluated at $t=0$ and $t=T$, respectively.
Analogously, the \textit{constrained expression} for the $\theta$ and $\Theta$ variables can be derived as
\begin{equation}\label{eq:cetheta}
	\begin{aligned}
		\theta(t) &= g_\theta(t) + \theta_f - g_\theta(T), \\
		\Theta(t) &= g_\Theta(t) + \Theta_0 - g_\Theta(0),         
	\end{aligned}
\end{equation}
where $g_\theta(t)$ and $g_\Theta(t)$ are the free functions for the $\theta$ and $\Theta$ coordinates, respectively.

The free functions are written as
\begin{equation}\label{eq:ff}
	\left\{\begin{aligned}
		g_r (t) &= \ds\sum_{j = 3}^m \xi_j \, h_j (\tau), \\
		g_\theta (t) &= \ds\sum_{j = 1}^m \zeta_j \, h_j (\tau),
	\end{aligned}\right.
	\quad
	\left\{\begin{aligned}
		g_R (t) &= \ds\sum_{j = 3}^m \xi^\prime_j \, h_j (\tau), \\
		g_\Theta (t) &= \ds\sum_{j = 1}^m \zeta^\prime_j \, h_j (\tau),
	\end{aligned}\right.
\end{equation}
where $\xi_j$, $\xi^\prime_j$, $\zeta_j$, and $\zeta^\prime_j$ are unknown coefficients, $h_j$ are the orthogonal Chebyshev polynomials (e.g. \cite{abramowitzstegun1972}), and $m$ is its highest truncation order. It is important to note that the orthogonal Chebyshev polynomials are valid only in the interval $-1\le \tau \le 1$, hence the change of the independent variable from $t$ to $\tau$ given by
$\tau=2t/T-1$ 
must be performed.
The sum starts at $j=3$ for $g_r$ and $g_R$ and at $j=1$ for $g_\Theta$ and $g_\theta$ because the terms of the polynomials must be linearly independent with the support functions, and there are three support functions in the radial direction and only one support function for the angular constraint.

The free functions of Eq.~(\ref{eq:ff}) are used in Eqs.~(\ref{eq:cer}) and (\ref{eq:cetheta}) to obtain new expressions of $r$ and $\theta$ used in Eq. (\ref{eq:motion_earth}).
It ensues in a set of two unconstrained differential equations that is discretized 
using the Chebyshev–Gauss-Lobatto nodes \cite{lanczos1988applied} distributed between 0 and $T$ with
\begin{equation}\label{eq:distr}
	t_k = \bigg(1- \cos \bigg(\frac{k \pi}{ N} \bigg) \bigg) \frac{T}{2}, \quad \text{for}~k=0,1,2,...,N,
\end{equation}
where $(N+1)$ is the number of equations generated by this procedure for each equation in Eq.~(\ref{eq:motion_earth}). These equations are then solved for the unknown coefficients $\xi_j$ for $j=3,4,...,m$ and $\zeta_j$ for $j=1,2,...,m$ using the nonlinear least squares optimization method \cite{M-ToC} to minimize the sum of the squares of the left sides of Eq.~(\ref{eq:motion_earth}) under the distribution given by Eq.~(\ref{eq:distr}). Further details on the TFC procedure can be seen in \cite{U-ToC,M-ToC}.

\section{Results}
\label{sec:results}

\begin{figure}
	\centering
	\includegraphics[scale=0.4]{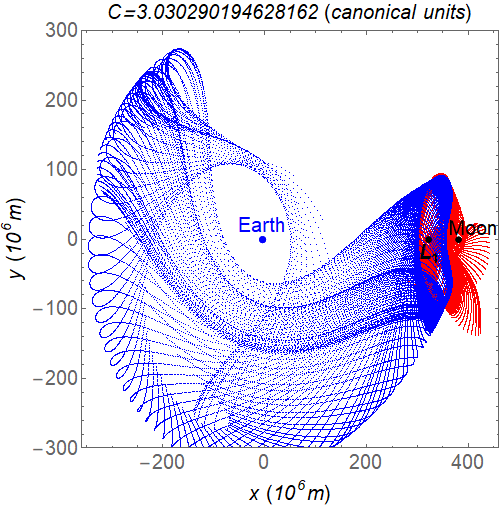}
	\includegraphics[scale=0.4]{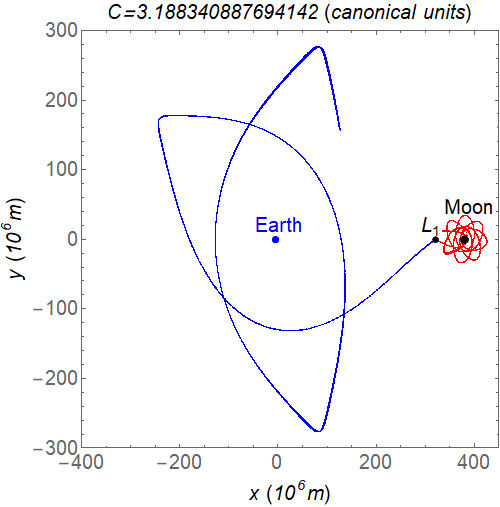}
	\caption{Stable (in blue) and unstable (in red) invariant manifolds associated to the Lyapunov orbits around $L_1$ with Jacobi constants equal to 3.030290194628162 and 3.188340887694142 in the left and right, respectively, in canonical units.
	}
	\label{fig:mani}
\end{figure}

The parameters adopted in this work are equal to those used to obtain the costs in the cislunar space in \cite{simo1995book,Yagasaki2004,topputo2013optimal,fastTFC}, and are shown in Table \ref{tab:parameters}.
\begin{table}[ht]
	\centering
	\begin{tabular}{cc}
		\hline
		$\mu_e$&$3.975837768911438\times10^{14}~\text{m}^3/\text{s}^2$ \\[0.5ex]
		$\mu_m$&$4.890329364450684\times10^{12}~\text{m}^3/\text{s}^2$ \\[0.5ex]
		$L$&$3.84405000\times10^{8}~\text{m}$\\[0.5ex]
		Radius of the Earth&$6.378\times 10^{6}~\text{m}$ \\[0.5ex]
		Radius of the Moon&$1.738\times 10^{6}~\text{m}$ \\[0.5ex]
		\hline
	\end{tabular}
	\caption{Values of the parameters used in this research.}
	\label{tab:parameters}
\end{table}
The Jacobi constant defines the Lyapunov orbit, and consequently it defines the invariant manifolds associated with this orbit. Although the dynamics around these structures is chaotic, the stable/unstable manifolds emanating from the Lyapunov orbits are well-defined two-dimensional surfaces in the system's phase space. In order to better understand them, we depict in Fig. \ref{fig:mani} stable (blue) and unstable (red) manifolds for the particular cases where the Jacobi constant is equal to 3.030290194628162 and 3.188340887694142, respectively.

The backward integration of a given point in the Lyapunov orbit perturbed in a stable direction generates a piece of a stable branch of the manifold, while the forward propagation of a point in the Lyapunov orbit perturbed in an unstable direction generates a piece of an unstable branch.
The union of all pieces for a given branch composes a surface and any point located on this surface can be used to perform the transfer. Focusing on the first leg of the journey, for example, each one of the points on the stable manifold has an associated total cost ($\Delta V$) regarding a bi-impulsive maneuver to leave the Earth and enter the manifold. The spacecraft is then transported with no extra costs to the Lyapunov orbit via the system's natural dynamics. We have then four free parameters for the transfer: the Jacobi constant $C$, the time of flight $T$, and the two coordinates which define the location of the point on the invariant manifold used for the transfer. These four parameters can be varied in order to find the best $\Delta V$. The same situation holds for the second leg of the journey, which comprises a maneuver from the unstable manifold to an orbit around the Moon.

\subsection{Transfers from Earth to Lyapunov orbit around $L_1$}
\label{sec:transf-LyL1}

We took advantage of the very efficient procedure shown in Sec. \ref{sec:models} to evaluate the costs associated with transfers for four values of the Jacobi constant: 2.94, 3.030290194628162, 3.091057859269563, and 3.188340887694142 (in canonical units). The length and time in canonical units are defined such that $L=1$ and $\omega=1$. For each value of the Jacobi constant, we chose 200 perturbed points on the Lyapunov orbit and integrated these backward in time for a minimum of 20 days and a maximum of 50 days (such values reflect the time it takes for a point on the manifold to reach the Lyapunov orbit). We later selected points inside the given time interval using time steps of 0.1 day. This process rendered us a total of 60000 points along the manifold surface. We then evaluated transfers from the Earth to each one of these points for times of flight varying from 0.1 to 10 days in intervals of 0.1 day. In total, we evaluated 24 million different trajectories to transfer a spacecraft from the Earth to the stable manifolds. The best costs for each Jacobi constant are shown in Table \ref{tab:jacobicosts}.
\begin{table}
	\resizebox{\textwidth}{!}{
		\begin{tabular}{c|c|c|c|c}
			\hline\noalign{\smallskip}
			Jacobi constant (canonical units) & 3.188340887694142 & 3.091057859269563 & 3.030290194628162  & 2.94   \\
			\noalign{\smallskip}\hline\noalign{\smallskip}
			$\Delta V$ (m/s) &  3758.66 & 3551.5 & 3438.5 & 3434.28\\
			\hline
	\end{tabular}}
	\caption{Minimal costs in terms of $\Delta V$ to transfer a spacecraft from the Earth to the stable manifold associated with the Lyapunov orbit for several values of the Jacobi constant $C$ evaluated from an ensemble of 24 million transfers.}
	\label{tab:jacobicosts}
\end{table}

The stable manifold has two branches, one that starts at the Earth's realm (the left branch, following our definition that Earth is located to the left of the Lyapunov orbit in the rotating reference frame), and one that starts at the Moon's realm (the right branch). The results showed thus far are related to transfers using the left branch. However, we performed a similar analysis using the right branch and obtained better results, which we now detail. It is somewhat counter-intuitive that designing transfers from an orbit around the Earth to the right branch of the stable manifold is more cost effective than using the left branch, given that these are farther away. The reason why this happens is because the right branch of the stable manifold gets very close to the Moon. This close encounter can reduce the transfer costs since the second impulse may be applied close to a local periselene (as seen from a local two-body approximation), near the surface of the Moon.

\begin{figure}
	\centering
	\includegraphics[scale=0.59]{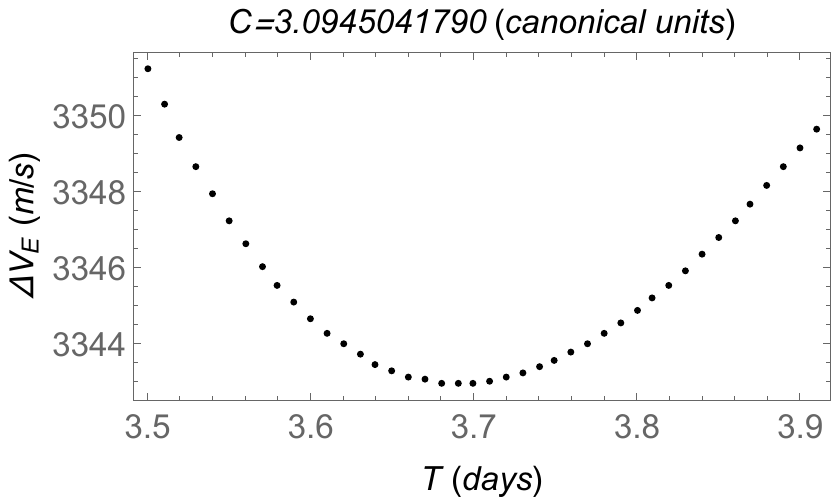}
	\caption{Costs to transfer from the Earth to the invariant manifolds associated to the Lyapunov orbits for the Jacobi constant $C=3.0945041790$ in canonical units. The transfer is done using a bi-impulsive maneuver with a swing-by with the Moon.
	}
	\label{fig:cfixed}
\end{figure}

We followed the same method used for the left branch, but this time we refined our results even further by looking at finer scales on the parameters. We started with a wide range for the parameter $C$ and for the location on the manifold, and we numerically investigated the costs to enter in each one of these points for a time of flight $T$ varying from 0.1 day to 10 days, in a total of 1397500 evaluated transfers, a procedure which led to a minimum. We repeated this procedure with other configurations in order to avoid local minima, and we found a global minimum at $C=3.0945041790$. We then fixed this value for the Jacobi constant, and evaluated the costs for 194825 points on the manifold for times of flight between 3.0 and 5.0 days in steps of 0.1 day, for a total of 4091325 transfers.
Later, we refined our results further by varying the time of flight $T$ from 3.5 days to 3.9 days in steps of 0.01 days, evaluating $7241960$ more transfers, and filtered out those that crashed with the surface of the Moon. The results for the minima are shown in of Fig. \ref{fig:cfixed}. The global minimum found is $3342.96$ m/s for a time of flight of $T=3.69$ days. 

\begin{figure}
	\centering
	\includegraphics[scale=0.6]{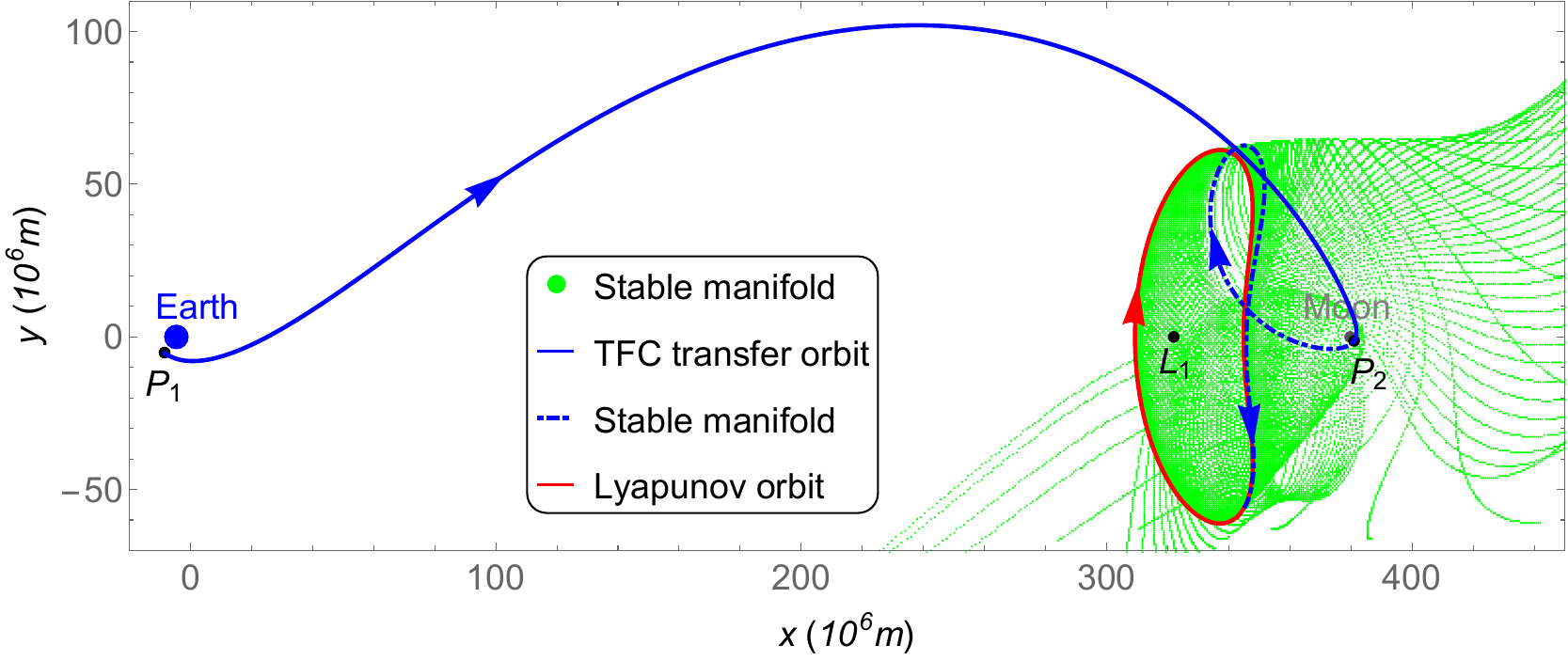}
	\caption{Orbit transfer from the Earth to the Lyapunov with a Jacobi constant 3.0945041790. The magnitude of the impulses applied at points $P_1$ and $P_2$ are 3142.09 m/s and 200.87 m/s, respectively, for a total $\Delta V_E$ of 3342.96 m/s, and the time of flight for the transfer is 3.69 days. The green trajectories form the right branch of the stable manifold and the dashed blue orbit represents the continuation of the stable manifold from $P_2$ up to 10 days, illustrating its connection with the Lyapunov orbit. The closest distance from the transfer orbit found via TFC to the surface of the Moon is 78 km, while the second impulse is applied in $P_2$ at an altitude of 142 km. The arrows indicate the directions of the motion along the orbits.}
	\label{fig:orbit1}
\end{figure}

The orbit associated with the lowest cost is depicted in Fig. \ref{fig:orbit1} in the rotating frame of reference centered at the barycenter, while its data (position and velocity) are disclosed in Table \ref{tab:points}. This orbit transfers a spacecraft from the point $P_1$ to the point $P_2$ and was generated via TFC by the procedure detailed in this paper. In fact, the velocity at $P_1$ must be tangent to the orbit, according to the constraint $\dot{r}(0)=0$ shown in Table \ref{tab:constraints}. The point $P_2$ is located at an altitude of 142.26 km (measured from the surface of the Moon), while the minimum distance of the spacecraft from the surface of the Moon during the transfer is 72.87 km. After the impulse applied at $P_2$, the spacecraft is transported to the Lyapunov orbit through the natural dynamics inherent to the CRTBP, although in reality a small control is required to compensate for perturbations not considered in this work. The found transfer in blue takes 3.69 days to complete, while the dashed blue orbit along the invariant manifold represents the 10 days flight needed to rejoin the Lyapunov orbit from the point P2. After that, the spacecraft can be easily transferred to the Lyapunov orbit, where it can stay orbiting the Lagrangian point $L_1$ using a thrust control system to compensate for other perturbations. As desired, the spacecraft can either make the travel back to the Earth using a swing-by maneuver with the Moon in a reverse procedure or it can be transferred to the Moon using a similar strategy given by the natural dynamics of the CRTBP combined with the TFC transfer orbit, as will be done in the next section.
\begin{table}
	\resizebox{\textwidth}{!}{
		\begin{tabular}{c|c|c|c|r}
			& Position coordinates $(x,y)$  $\times 10^6$ m & Velocity vector $(\dot{x},\dot{y})$ before impulse (m/s) & Velocity vector $(\dot{x},\dot{y})$ after impulse (m/s) & $\Delta V$ (m/s)\\
			\hline  
			$P_1$ & (-8.44265550170148, -5.348828138546912) & (6355.308406789542,-4481.625958616931) & (8923.14451331409,-6292.408412569893) & 3142.09 \\
			$P_2$ & (381.00217674797364, -1.388400819981988) & (-1434.18433217481,-1983.39118640976) & (-1301.6775762958073,-1832.4178605997574) & 200.87 \\
			$P_3$ & (357.9175878955692, -55.569866112517055) & (8.878774619088471,-145.19417907514665) & (9.25312317424971,-144.4748800109339) & 0.81 \\
			$P_4$ & (381.2553054029977,1.031770494048325) & (-1276.577060988742,1881.988040738359) & (-912.912844015338,1345.857690206811) & 647.83 \\
			\hline
	\end{tabular}}
	\caption{Points $P_1$, $P_2$, $P_3$, and $P_4$ and velocities for an spacecraft arriving and leaving these points in the rotating frame centered at the barycenter. The cost $\Delta V$ is the magnitude of the difference of the velocities before and after the impulse.}
	\label{tab:points}
\end{table}

The fuel equivalent cost $\Delta V$ to perform the transfer from the Earth to the Lyapunov orbit found in this work is the sum of the magnitudes of the specific impulses applied at $P_1$ and $P_2$, which are 3142.09 m/s and 200.87 m/s, respectively. Thus, the $\Delta V$ to perform a transfer from a circular parking orbit of altitude 167 km around the Earth to the Lyapunov orbit around $L_1$ is 3342.96 m/s.
\begin{table}
	\centering
		\begin{tabular}{c|c|c|c|c|c}
			alt (km)& Destination & $\Delta V$ (m/s) & ToF (days) &  NoI & Reference \\ 
			\hline
			167 & $L_1$ Lyapunov orbit & 3342.96 & 13.69 days & 2 &  This work  \\         
			185 & $L_1$ Lyapunov orbit & 3391.8823 & 16.9050 days & 2 &  \cite{QI2019180} \\ 
			185 & $L_1$ Halo orbit & 3589.7 & 22.1 days & 2 &  \cite{Parker2008} \\ 
			185 & $L_1$ Lyapunov orbit & 3596.7871 & 26.1434 days & 2 & \cite{QI2019180}  \\
			200 & $L_1$ Halo orbit & 3659.5 & 4.67 days & 2 &  \cite{rauschmaster}  \\
			200 & $L_1$ Halo orbit & 3660 & 4.1 days & 2 &  \cite{folta13} \\ 
			200 & $L_1$ Halo orbit & 3682.4 & 4.95 days & 2 &  \cite{rauschmaster}  \\
			200 & $L_1$ Halo orbit & 3802.46 & 12.18 days & 2 &  \cite{rauschmaster}  \\
			200 & $L_1$ Halo orbit & 3359.32 & 22.27 days & 3 &  \cite{Zeng2016} \\ 
			200 & $L_1$ Halo orbit & 3368.61 & 21.70 days & 3 &  \cite{Zeng2016} \\ 
			200 & $L_1$ Lagrangian point & 3439.8 & 22.9 days & 3 &  \cite{Li2010} \\ 
			200 & $L_1$ Halo orbit & 3513 & 7.81 days & 3 &  \cite{folta13} \\ 
			\hline
		\end{tabular}
	\caption{Transfers from a parking orbit around the Earth with altitude ``alt'' to orbits around $L_1$ evaluated under the CRTBP model. NoI is the number of impulses required to perform the transfer. ToF is the time of flight, with different definitions among the references - it can be seen as an estimative for mission design.}
	\label{tab:comparison_L1}
\end{table}
The costs involved in transfers from a parking orbit around the Earth to periodic orbits around Earth-Moon Lagrangian $L_1$ are shown in Table \ref{tab:comparison_L1} for several works available in the literature.
In particular, a comprehensive analysis on transfer from a parking orbit around the Earth to Lyapunov orbits can be seen in \cite{QI2019180}, including the one shown in the line of Table \ref{tab:comparison_L1} associated with the cost value 3391.8823 m/s. It represents the most efficient (lowest cost and lowest ToF) transfer for short duration (less than 50 days) between the Earth and Lyapunov orbits around $L_1$ available in the literature \cite{QI2019180}. This transfer is similar to the best one selected in the present work shown in the first line of Table \ref{tab:comparison_L1}, with a flyby around the Moon. 
However, in \cite{QI2019180}, the transfer stars from an initial orbit around the Earth of altitude 185 km, whereas this altitude is 167 km in the present work.
Using the Jacobi constant, we estimate that the minimum cost needed to transfer a spacecraft from 167 km to 180 km altitude is 10.68 m/s.
The details of a similar computation are presented in Sec. \ref{sec:complete_transfer}.
Hence, although these transfers are similar - due to the fact that the second impulse is applied during a close encounter with the Moon in a retrograde direction - the transfer proposed in the current work saves 59.60 m/s.


\subsection{Transfers from the Lyapunov orbit around $L_1$ to the Moon}
\label{sec:transf-L1M}


In this Section, we consider the problem of transferring a spacecraft from the unstable manifold to the Moon using a bi-impulsive maneuver. The Jacobi constant is set at $C=3.0945041790$, which is related to the lowest cost $\Delta V$ found for the first leg of the journey in Sec. \ref{sec:transf-LyL1}, namely, transferring the spacecraft from the Earth to the stable manifold. 
Here we must find a point $P_3$ along the unstable manifold for which a low-cost impulse may be applied such that the spacecraft is transferred to a circular orbit of 100 km of altitude around the Moon with the aid of another impulse applied at a point named as $P_4$, whose coordinates are also to be found using the procedure based on TFC proposed in this paper. Since the final radius around the Moon is constrained at $R_f=(1738+100)$ km (radius of the Moon plus 100 km altitude), the angle $\Theta$ is obtained by our optimization procedure taking into consideration the constraints given in Table \ref{tab:constraints}. This means that the final velocity at $P_4$ must be tangential to the circular orbit around the Moon, and then the condition $\dot{R}=0$ must be satisfied.

\begin{figure}
	\centering
	\includegraphics[scale=0.6]{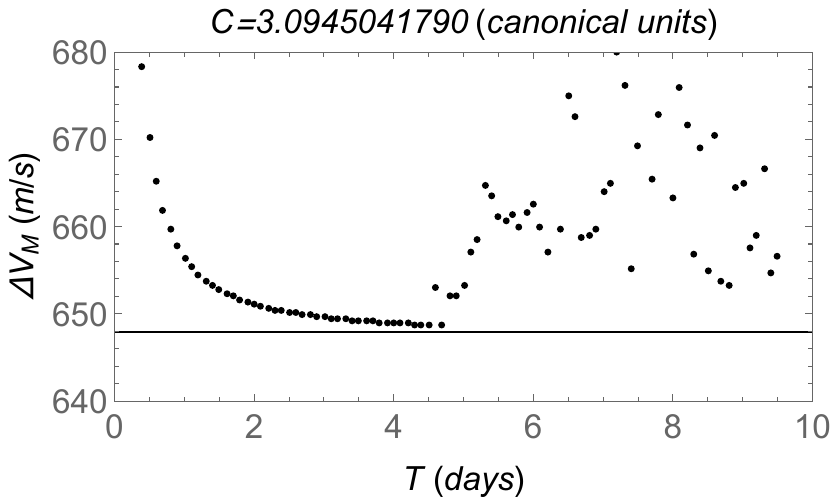}
	\caption{Minimum cost $\Delta V_M$ to transfer from the unstable manifold to the Moon as a function of time of flight. The manifold is associated to the Lyapunov orbit with Jacobi constant $C=3.0945041790$. The lowest $\Delta V_M$ cost is 0.648644 km/s for a time of flight of 4.7 days. This value is obtained from an ensemble of one million transfers. The horizontal line at $\Delta V_M=647.877$ m/s is the theoretical minimum defined in Sec. \ref{sec:comparison}.}
	\label{fig:costs_leave}
\end{figure}
\begin{figure}
	\centering
	\includegraphics[scale=0.5]{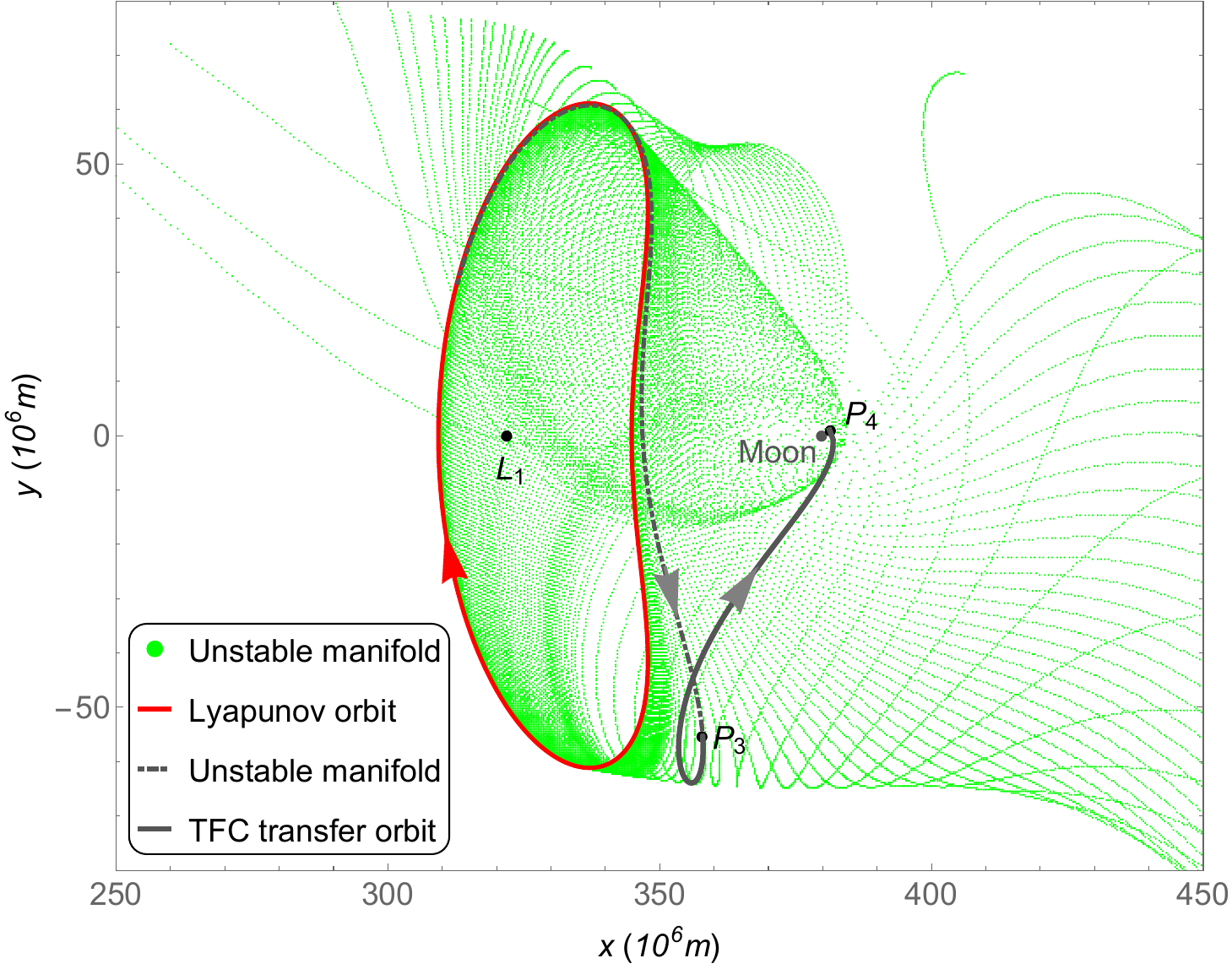}
	\caption{Lyapunov orbit to Moon transfer in the rotating frame. The magnitude of the impulses applied at points $P_3$ and $P_4$ are 0.81 m/s and 647.83 m/s, respectively, for a total $\Delta V_M$ of 648.64 m/s. The time of flight of the TFC transfer orbit represented by the continuous gray curve is 4.7 days, while the dashed line corresponds to the 8 days flight in the invariant manifold. The impulse at $P_4$ is applied to circularize the orbit at an altitude of 100 km. The arrows indicate the direction of motion along the orbits.}
	\label{fig:orbit2}
\end{figure}

In order to find the best transfer orbit, we chose 10000 points on the unstable manifold and we considered a time of flight $T$ in the interval between 0.1 and 10 days with a step of 0.1 day. The lowest fuel equivalent cost $\Delta V_M$ of the bi-impulsive maneuver to design this transfer is shown in Fig. \ref{fig:costs_leave}.
The lowest total cost found is 648.644 m/s corresponding to a time of flight of 4.7 days. The transfer orbit associated with this lowest value is shown in the continuous gray curve in Fig. \ref{fig:orbit2} between points $P_3$ and $P_4$. Such points and the transfer orbit between them were obtained using the procedure detailed in this paper. 
The gray dashed curve shown in Fig. \ref{fig:orbit2} represents the 8 days flight on the unstable manifold to reach the point $P_3$ from the Lyapunov orbit.

\subsection{Complete transfer from the Earth to the Moon using manifolds}
\label{sec:complete_transfer}

The results obtained in Secs. \ref{sec:transf-LyL1} and \ref{sec:transf-L1M} can be used to design a more complete mission involving a transfer from the Earth to the Moon combined with a close encounter with the Moon 3.69 days after leaving Earth
and a visit to the Lyapunov orbit, where it could stay using a flight control, depending on the objectives of the mission. Four impulses are applied at points $P_1$, $P_2$, $P_3$, and $P_4$ to perform this transfer. A first impulse is applied around the Earth at $P_1$ to start the transfer. The second impulse is applied at $P_2$ to insert the spacecraft in the stable manifold in a combination with a gravity-assist maneuver close to the Moon.
The spacecraft can then use the natural dynamics inherent to the CRTBP to be transported in the direction of the Lyapunov orbit through this stable manifold.
Although the time of flight to reach the Lyapunov orbit using the manifold may be large under the CRTBP, after about 10 days the spacecraft is less than 100 km close to the Lyapunov orbit, as can be seen at the end of the dashed blue line in Fig. \ref{fig:orbit1}. In this sense, a control can be used to adjust this time. For instance, the dashed gray line in Fig. \ref{fig:orbit2} connects the Lyapunov orbit with point $P_3$. In order to show that a control can be used to adjust the times of flight, we evaluated transfers between the last point of the blue dashed curve and the first point of the gray dashed curve. The stable and unstable manifolds can be connected through an orbit transfer ``surrounding'' the Lyapunov orbit in 5.4882 days using a bi-impulsive maneuver costing 0.4263 m/s, which should be performed thrust a control system. This Lyapunov orbit has a period of about 13.75 days, thus the spacecraft could stay in this orbit around $L_1$ for periods multiple of this value, depending on the objectives of the mission. 
After that, the spacecraft ``leaves'' the Lyapunov orbit through the dashed gray curve, ending at point $P_3$ in 8 more days.
At this point, a very small impulse with magnitude 0.81 m/s is applied such that the spacecraft enters in the transfer orbit to the point $P_4$, which is obtained through the procedure shown in Sec. \ref{sec:models} using Moon-centered polar coordinates. This cost is close to zero because the TFC transfer orbit is adjacent to the manifolds (see Fig. \ref{fig:orbit2}). Afterwards, a final impulse is then applied at point $P_4$ in the tangential direction (according to the constraint $\dot{R}=0$ shown in Table \ref{tab:constraints}) 4.7 days after the spacecraft passes point $P_3$, to circularize the orbit around the Moon at an altitude of 100 km.

\begin{figure}
	\centering
	\includegraphics[scale=0.5]{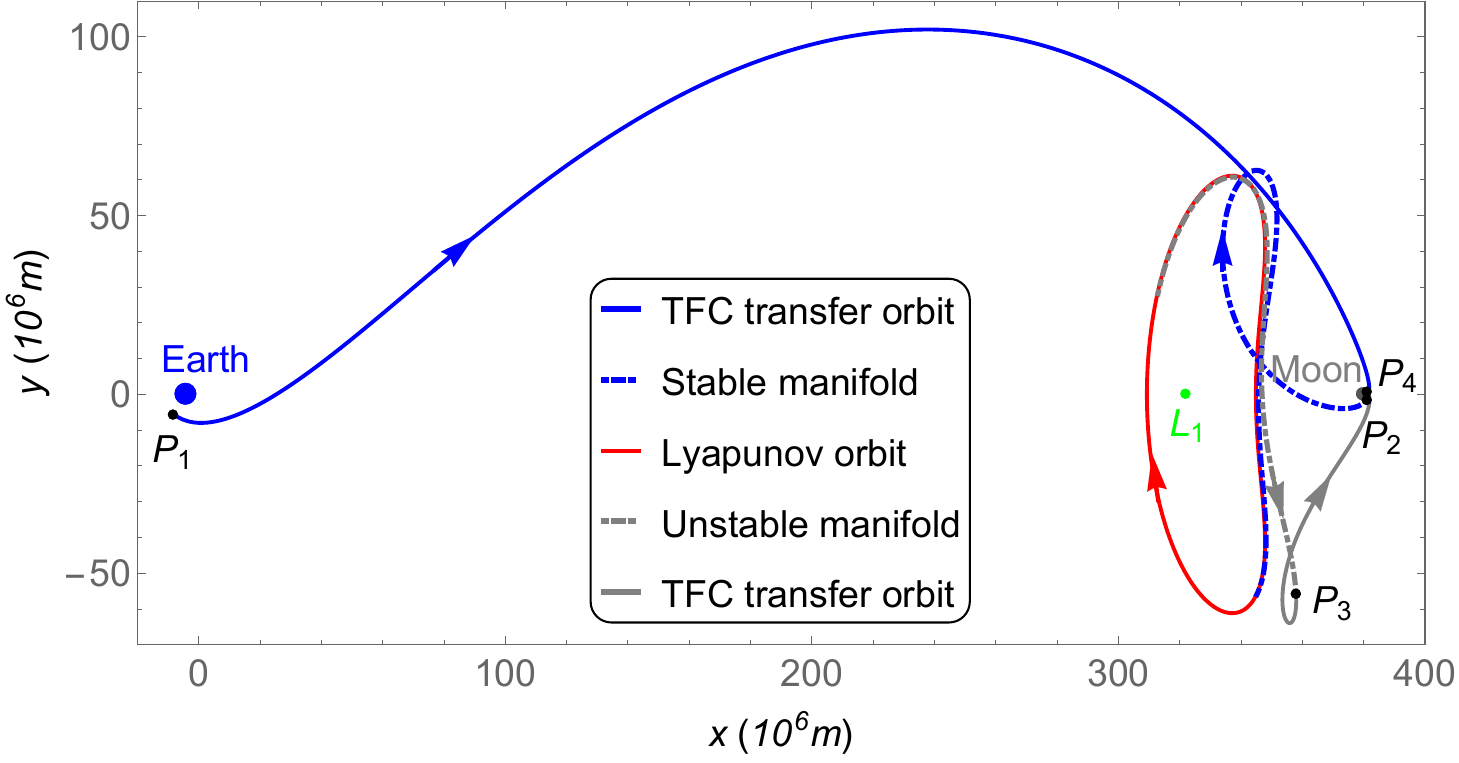}
	\includegraphics[scale=0.5]{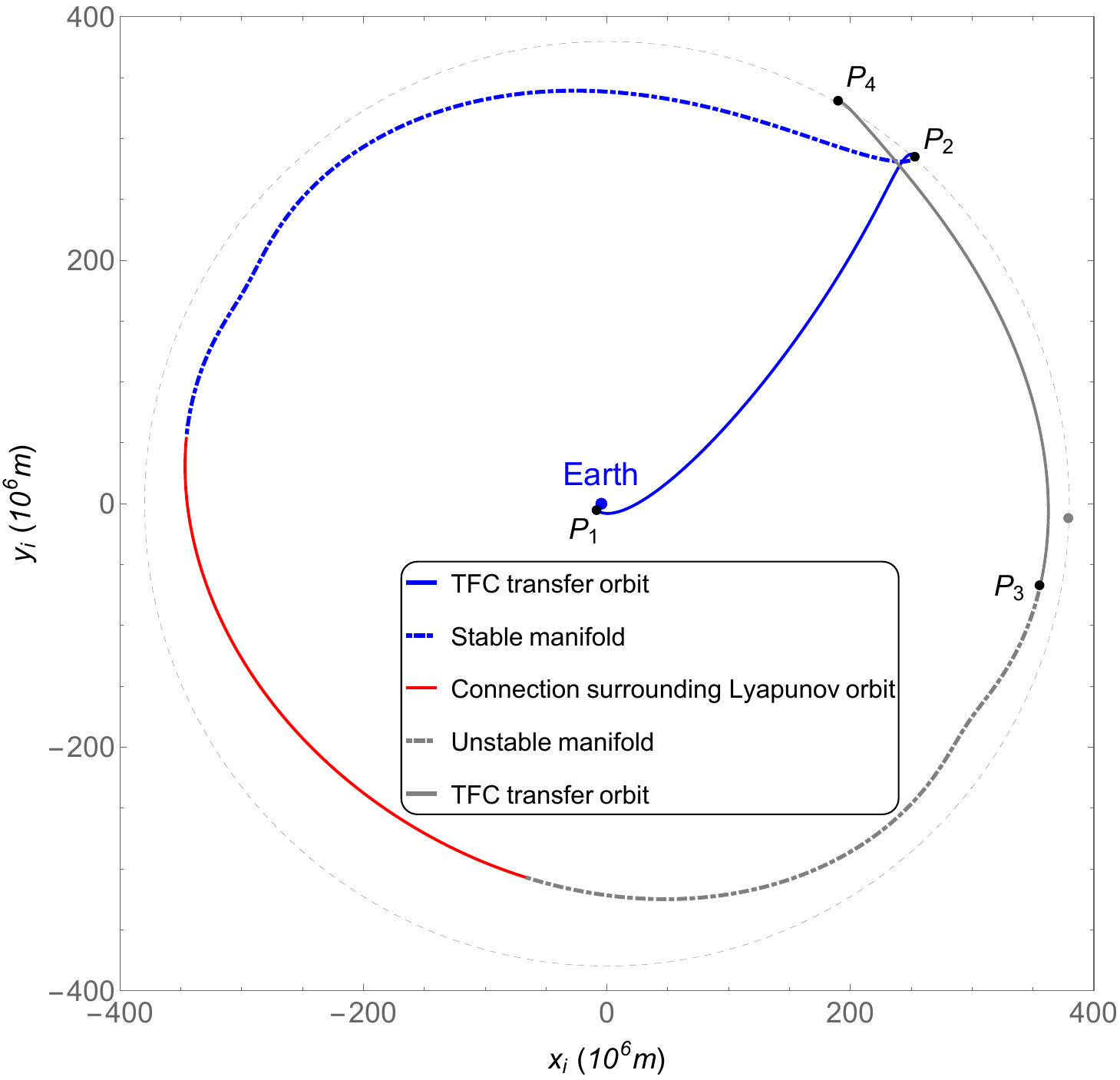}
	\caption{Earth to Moon transfer via a Lyapunov orbit around $L_1$ in the rotating (top) and inertial (bottom) frames. The magnitude of the impulses applied at points $P_1$, $P_2$, $P_3$, and $P_4$ are 3142.09 m/s, 200.87 m/s, 0.81 m/s, and 647.83 m/s, respectively, for a total of 3991.60 m/s. The time of flight of the blue and gray TFC transfer orbits are 3.69 days and 4.7 days, respectively, while the blue and gray dashed orbit in the invariant manifold  are shown for 10 and 8 days, respectively. The spacecraft may stay in the Lyapunov orbit (shown in red) as long as required by the mission using control to compensate for perturbations not considered in this work. The arrows indicate the directions of the motion along the orbits. The gray dashed thin curve represents the orbit of the Moon in the inertial frame, whose position is shown as the gray dot for when the spacecraft is at $P_3$.}
	\label{fig:orbit3}
\end{figure}

The orbits involved in the complete Earth-to-Moon transfer are shown in Fig. \ref{fig:orbit3}, and the positions and velocities of the orbits at the four points $P_1$, $P_2$, $P_3$, and $P_4$ before and after the burns are shown in Table \ref{tab:points}. The complete transfer from the Earth to the Moon costs then 3991.60 m/s (plus fuel spent in control) and can be done in a total time of flight (using the control connection explained above) of 31.8782 days.
As explained above, this time can be extended in multiples of the period of the Lyapunov orbit (13.75 days), according to the mission.

\label{sec:comparison}


From the Jacobi constant, Eq. \eqref{eq:C}, it is possible to calculate the minimum cost needed to transfer the spacecraft between two trajectories with different energies. We can use such a value to estimate the minimum theoretical impulse, $\Delta V_{min}$, needed to perform the transfer described in Section \ref{sec:complete_transfer}.

First, we consider the transfer corresponding to the smallest $\Delta V$ to go from the initial orbit around the Earth to the manifold defined by the Jacobi constant $C_2=3.0945041790$.
This transfer starts at the point $P_1$
situated on the initial orbit, where the spacecraft has a circular motion with a tangential velocity $v_1$, and is associated to the cost $\Delta V_E=3342.96\,\mathrm{m/s}$.
The initial circular orbit around the Earth corresponds to the value of the Jacobi constant $C_1=58.30280824908883$ in canonical units.
We suppose that we apply an impulse $\Delta V_{Emin}$ at the initial point to change the value of the Jacobi constant from $C_1$ to $C_2$. This impulse is applied in the same direction of the initial velocity, which corresponds to the minimum value for the impulse, and must then satisfy
\begin{equation}
	\Delta V_{Emin}=\sqrt{\left(C_1-C_2\right)+v_1^2}-v_1.
\end{equation}
Considering the condition $C_2=3.0945041790$ in canonical units obtained in the previous section, we find the theoretical minimum $\Delta V_{Emin}=3099.02\,\mathrm{m/s}$. Thus, the minimum cost $\Delta V_E$ obtained in this research is $243.94\,\mathrm{m/s}$ higher than the theoretical minimum. It is possible to decrease this cost, but the transfer would require longer times of flight and, due to this reason, the analyses should consider other perturbations (other models than the CRTBP) in order to be useful for mission design, as to estimate/minimize transfers costs or to be used as initial guess in an ephemeris model. In this case, the analyses then would depend on the time epoch, since the perturbations (e.g. gravity of other bodies) are functions of time.
The results obtained in this work can be adopted as initial guess for the desired epoch in order to obtain a more accurate orbit in a high fidelity model.
In fact, extra fuel can be saved in an ephemeris model in case a specific epoch is adopted to perform the transfer. For instance, 63 m/s is saved using an ephemeris model in comparison with the CR3BP, in the case where the transfer performed in the epoch ``UTC 2018 FEB 15 00:00:00'' \cite{QI2019180}.


We now perform the same calculation for the transfer from the manifold associated to the Jacobi constant $C_2$ to the final orbit around the Moon.
This transfer ends at the point $P_4$,
where the spacecraft has a circular motion around the Moon with tangential velocity $v_2$, and is associated to the cost $\Delta V=648.644\,\mathrm{m/s}$.
The final orbit corresponds to a Jacobi constant of $C_3=5.508041392480298$, and the minimum value for the impulse satisfies
\begin{equation}
	\Delta V_{Mmin}=\sqrt{\left(C_3-C_2\right)+v_2^2}-v_2,
\end{equation}
for which we find $\Delta V_{Mmin}=647.877\,\mathrm{km/s}$. Thus, the minimum cost $\Delta V_M$ obtained in this research is only $0.767\,\mathrm{m/s}$ higher than the theoretical minimum. There is a lot of options with respect to the time of flight to design this transfer with $\Delta V_M$ close to the theoretical minimum, as shown in Fig. \ref{fig:costs_leave}. This is due to the fact that the manifolds approach the surface of the Moon. The costs to transfer from the Earth to the manifolds are much higher than the theoretical minimum because the manifolds do not approach the surface of the Earth in a short time.



The costs to transfer a spacecraft from the Earth to the Moon are obtained in several works available in the literature. Although they do not include the benefits of the transfer designed in this paper, such as a previous close encounter with the Moon and a visit to the Lyapunov orbit, a comparison of the costs to the complete Earth-Moon transfer case is done next. The time of flight of the complete Earth-Moon transfer obtained in this work is 31.8782 days, associated to a total cost of 3991.60 m/s. A transfer from the Earth to the Moon done in a time of flight of 31 days is associated with the cost given by 3924.94 m/s, evaluated under the restricted three-body problem in \cite{YAGASAKI2004313}. In fact, many authors evaluated Earth-to-Moon transfers costs using the R3BP, as can be seen in Table \ref{tab:EMcosts}.
This time of flight is similar to the one obtain in this research, although the cost is 66.66 m/s lower. Thus, we can conclude that an Earth-Moon transfer cost is increased by this amount in case an intermediary orbit is designed around the Lagrangian $L_1$ after a close encounter with the Moon, instead of using the orbit proposed by \cite{YAGASAKI2004313} (see Fig. 8g of this reference).
\begin{table*}[!ht]
	\centering
	\resizebox{\textwidth}{!}{
		\begin{tabular}{c|c|c|c|c|c|c|c|c|c}
			\hline
			\multicolumn{1}{c|}{$T$ (days)}&3.1 to 3.4    &  4.5625&  14&31&68&85& 193&255&292               \\
			\multicolumn{1}{c|}{$\Delta V$ (m/s)} &4042 to 4007&3946.93& 3950&3925&3920&3861&3899&3894&3824 \\
			\multicolumn{1}{c|}{Literature reference:}&\cite{doi:10.2514/1.7702}& \cite{fastTFC}& \cite{doi:10.2514/1.7702}&\cite{YAGASAKI2004313}&\cite{doi:10.2514/1.7702}&\cite{doi:10.2514/1.7702}&\cite{topputo2005earth}&\cite{topputo2005earth}&\cite{Pernicka1995}      \\
			\hline
	\end{tabular}}
	\caption{Best trade-offs between $\Delta V$ and times of flight found in the literature for impulsive transfers from the Earth to the Moon under the restricted three-body problem. The altitudes of the orbits around the Earth and the Moon are $167$ km and $100$ km, respectively.}
	\label{tab:EMcosts}
\end{table*}

\section*{Conclusions}

In this paper, we designed orbital transfers between the Earth and the stable invariant manifold associated to the Lyapunov orbit around $L_1$ and from the unstable invariant manifold associated to the same Lyapunov orbit to the Moon. These transfers are designed using the procedures detailed in this paper, based on TFC combined with strategic constraints, allowing us to solve the problem and generate the results with reasonably low computational resources. The lowest cost of the first part (between the Earth and the Lyapunov orbit) is associated with a gravity-assist maneuver with the Moon. This means that a multi-purpose mission can be designed using the results obtained in this paper: a first close (and fast) encounter with the Moon, a transfer to the Lyapunov orbit, and a final transfer to the Moon. The results shown in this paper can be useful for such a type of multi-purpose mission, involving several stages. For instance, a mother spacecraft can travel to the Lyapunov orbit, from where smaller satellites can be launched to the Moon and easily tracked. Using the procedures and results of this paper, a similar strategy can be adopted to design transfers from the Moon back to the mother spacecraft and also from the Lyapunov orbit to the Earth, depending on the objectives of the mission.

The renewed global interest in lunar exploration is increasing the traffic to the Moon and redundant monitoring will be required to ensure flight safety and collision avoidance, both of which can benefit from the orbital transfer designed here. The smaller orbital volume around the Moon indicates the need for innovative orbital approaches which may consider the Lyapunov orbits around the Lagrangian points in the Earth-Moon system. These orbital solutions present a considerable strategic interest for permanent lunar presence and are key to the development of new design techniques which accommodate extended spacecraft lifetime for scientific exploration and lunar remote sensing. In particular, the solutions explored here support spacecraft orbits that cover a large surface area of the Moon, thus enabling a wide remote sensing footprint and relay communication coverage to surface instruments, rovers or bases in the Moon surface for extended periods with reduced fuel costs. This is paramount to optimize new exploration concepts and improve mission system design.

\section*{Acknowledgements}
AKAJ, TV, and DM were supported by the project Centro de
Investiga\c{c}\~ao em Ci\^encias
Geo-Espaciais, reference UIDB/00190/2020, funded by COMPETE 2020 and
FCT, Portugal. VMO was partially supported by the São Paulo Research Foundation (FAPESP, Brazil), under Grants No. 2021/11306-0, 2022/12785-1, and by Fundação para a Ciência e a Tecnologia (FCT, Portugal), under projects UIDP/04564/2020\footnote{\href{https://doi.org/10.54499/UIDP/04564/2020}{https://doi.org/10.54499/UIDP/04564/2020}.} and UIDB/04564/2020\footnote{\href{https://doi.org/10.54499/UIDB/04564/2020}{https://doi.org/10.54499/UIDB/04564/2020}.}.

\section*{Contributions}

AKAJ and LBTS generated the results. AKAJ wrote its first version. VMO suggested the initial concept. DM provided the financial support. All authors analyzed the results and revised the manuscript.

\bibliography{ref2}



\end{document}